\documentclass[]{aa}

\usepackage{graphicx}
\usepackage{txfonts}
\usepackage{natbib}
\usepackage{hyperref}
\usepackage{xspace}

\newcommand{\kms}{km\,s$^{-1}$\xspace}
\newcommand{\Msun}{$M_\odot$\xspace}
\newcommand{\god}{\textsc{Godunov-MHD}\xspace}
\newcommand{\bfi}{$\langle B_\phi \rangle$\xspace}

\begin{document} 

\title{Magnetic field evolution in dwarf and Magellanic-type galaxies}
\subtitle{}

\author{H. Siejkowski\inst{1}
\and M. Soida\inst{2}
\and K.T. Chy\.zy\inst{2}}

\institute{
AGH University of Science and Technology, ACC Cyfronet AGH, ul. Nawojki 11, PO Box 386, 30-950, Krak\'{o}w 23, Poland,\\ \email{h.siejkowski@cyfronet.pl}
\and
Astronomical Observatory of the Jagiellonian University, ul. Orla 171, 30-244 Krak\'ow, Poland
}

\abstract
{}
{Low-mass galaxies radio observations show in many cases surprisingly high levels
of magnetic field. The mass and kinematics of such objects do not favour the
development of effective large-scale dynamo action. We attempted to check
if the cosmic-ray-driven dynamo can be responsible for measured magnetization
in this class of poorly investigated objects. We investigated how starburst events on the whole, as well as when part of the galactic disk, influence the magnetic field evolution.}
{We created a model of a dwarf/Magellanic-type galaxy described by gravitational potential constituted from two components: the stars and the dark-matter halo. The model is evolved by solving a three-dimensional (3D) magnetohydrodynamic equation with an additional cosmic-ray component, which is approximated as a fluid. The turbulence is generated in the system via supernova explosions manifested by the injection of cosmic-rays.}
{The cosmic-ray-driven dynamo works efficiently enough to amplify the magnetic field
even in low-mass dwarf/Magellanic-type galaxies. The $e$-folding times of magnetic energy growth are 0.50 and 0.25\,Gyr for the slow (50\,\kms) and fast (100\,\kms) rotators, respectively. The amplification is being suppressed as the system reaches the  equipartition level between kinetic, magnetic, and cosmic-ray energies.  An episode of star formation
burst amplifies the magnetic field but only for a short time while increased star formation activity holds. We find that a substantial amount of gas is expelled from the galactic disk, and that the starburst events increase the efficiency of this process.}
{} 

\keywords{
dynamo
--
galaxies: dwarf
--
galaxies: magnetic fields
--
galaxies: starburst
--
magnetohydrodynamics (MHD)
--
methods: numerical
}

\maketitle


\section{Introduction}
Dwarf galaxies are the most abundant galaxies in the universe but they are also difficult to observe due to their low luminosity and small size. 
Evolution of stars and the interstellar medium (ISM) in dwarfs can be studied in the absence of spiral density waves and on a small scale. 
Low masses and metallicities of dwarf galaxies suggest that they could have been the building blocks of larger galaxies, constituting the key element in the theory of 
hierarchical galaxy formation. 

There is no clear break to distinguish a dwarf from a larger (irregular) galaxy of the Magellanic type
but according to working definitions, all galaxies fainter than $M_v=-17$ are regarded as dwarfs 
\citep{tolstoy}. The Small Magellanic Cloud is close to 
this boundary and resembles some larger dwarfs from the Local Group, such as IC\,1613 and NGC\,6822, 
while the Large Magellanic Cloud is more like low-luminosity spirals; for example, M\,33.

In massive spiral galaxies, the ISM involves sufficiently 
strong shearing motions due to differential rotation and the Coriolis force
to generate large-scale magnetic fields by dynamo processes \citep{widrow}.
Such magnetic fields have been discovered in a number of spiral galaxies and investigated 
through synchrotron emission and Faraday rotation effects \citep{beck16}.
The low-mass, dwarf/irregular galaxies are also rich in gas but reveal much more 
chaotic gas motions and slow rotation, which might not provide favourable conditions for generating 
strong and large-scale magnetic fields. 

In spite of these considerations, strong magnetic fields ($>10\,\mu$G) were actually observed in 
the dwarf galaxies IC\,10 \citep{chyzy16} and NGC\,1569 \citep{kepley10}, as well as even a regular field 
in NGC\,4449 \citep{chyzy00}. However, a systematic survey of galaxies in the Local
Group \citep{chyzy11} led to radio detection at 2.64\,GHz of only 3 out of 12 dwarfs. 
The results obtained indicated that the mean strength of magnetic fields  
in the local dwarf galaxies is small, about $4\,\mu$G, and related to the galactic 
star formation rate (SFR). 

The generation and evolution of magnetic fields in dwarf galaxies were simulated by a cosmic-ray (CR)-driven dynamo in the shearing-box approximation \citep{siejkowski10} as well as in the global model \citep{siejkowski14}. 
In this model, the CRs and magnetic seed fields are injected into the ISM by randomly exploding supernovae (SN). 
The simulations prove that even in the slowly rotating dwarf galaxies this kind of dynamo is able to amplify magnetic 
field with an exponential growth rate, reproducing the observed magnetic field strengths. The $e$-folding time 
of magnetic azimuthal flux growth appeared to be correlated with the initial rotation speed and was rather long (about 1\,Gyr) 
for a galaxy rotating with a speed $v_\phi=40\,$km\,s$^{-1}$. The simulations also show a significant loss of magnetic 
field from the simulated domain \citep{siejkowski10}. Such an effect depends on the SN rate and is due to gas 
escaping through galactic winds driven by cosmic rays \citep[see also][]{hanasz13}.
 
In these simulations, the production of stars was assumed to be the same throughout the evolution of galaxies. 
However, this simplification may be too strong. Investigation of star formation history (SFH) of dwarf galaxies indicates 
that in fact their rates of star formation are quite diverse and show fluctuations in time \citep{weisz}. 
According to numerical simulations, the tidal-driven changes in star formation are rather short-lived, with a typical 
duration of the activity of about $10^8$\,yrs \citep{matteo}. Strong starbursts during which the star formation rate 
is increased by a factor larger than 5 are rare. IC\,1613 is a good example of a dwarf that calmly produces stars over 
its entire lifetime, while showing also some fluctuations \citep{skillman}. The galaxy has a small H\textsc{i} mass 
of $8\times 10^7$ and a stellar mass of $1.1\times 10^8$  \citep{woo}. The current SFR is 0.003 \Msun\,yr$^{-1}$ \citep{mateo}, 
but the SFH reveals some episodes in the galaxy life with the SFR enhancement of about threefold \citep{skillman}. 

Another aspect of evolution of low-mass galaxies not addressed in numerical experiments to date is the often 
observed irregular distribution of star-forming regions in galactic disks. Such an asymmetric distribution is evident 
in for example IC\,10, dominated by a giant H\textsc{ii} region in its southern part. The observed morphology and the strength of 
magnetic field are strongly related to the distribution of H\textsc{ii} regions within the galactic disk \citep{chyzy16}.

So as to address the above issues, in this paper a series of 3D simulations of the CR-driven dynamo 
for low-mass galaxies are performed. In the first place, we simulate a calm evolution of galaxies at different speeds 
of rotation and different (but constant in time) levels of star formation activity. Below, we introduce a starburst 
(a rapid enhancement of star formation)  and study its impact on magnetic field, while finally investigating how a localised 
starburst, restricted to just part of the galactic disk, changes the magnetic field topology, and how this is related to 
the topology for the global starburst.
 
\section{Numerical model and setup}
\subsection{Cosmic-ray-driven dynamo}

The CR-driven dynamo model in dwarf/irregular galaxies has been implemented into a numerical code called \god \citep{kowal09}. The code solves the system of resistive magnetohydrodynamic (MHD) equations in a conservative form in 3D space \citep[e.g.][]{principles_mhd}:

\begin{align}
  \frac{\partial \rho}{\partial t} + \nabla \cdot (\rho \vec{v}) & = 0 \\
  \frac{\partial \rho \vec{v}}{\partial t} + \nabla \left[ \rho\vec{v}\vec{v}
  + \left(p + p_\mathrm{cr} + \frac{B^2}{8\pi}\right)\vec{I} - \frac{\vec{B} \vec{B}}{4\pi} \right]
  & =  -\rho \nabla \phi \\
  \frac{\partial \vec{A}}{\partial t} - \vec{v} \times (\nabla \times \vec{A}) - \eta \nabla \times
  (\nabla \times \vec{A}) & = 0,
  \label{eq:induction}
\end{align}

\noindent
where $\rho$, $p,$ and $\vec{v}$ are the gas density, pressure, and velocity, respectively, $p_\mathrm{cr}$ is the cosmic-ray pressure, $\vec{A}$ is the vector potential, $\vec{B} \equiv \nabla \times \vec{A}$ is the magnetic field, $\phi$ is the gravitational potential, $\eta$ is the magnetic turbulent diffusivity and $\vec{I}$ is the identity matrix. To close the system an isothermal equation of state is assumed:

\begin{equation}
  p \equiv \frac{c_s^2}{\gamma} \rho
  \label{eq:eos}
,\end{equation}

\noindent
where $c_s$ is the isothermal speed of sound and $\gamma$ is the adiabatic index of the gas.

Cosmic rays form a key component of the model. In our case they are approximated as a fluid, whose evolution is described by the diffusion-advection transport equation following \cite{schlickeiser_lerche}:

\begin{equation}
  \frac{\partial e_\mathrm{cr}}{\partial t} + \nabla(e_\mathrm{cr} \vec{v}) = \nabla (\hat{K} \nabla
  e_\mathrm{cr})
  - p_\mathrm{cr} (\nabla \cdot \vec{v}) + Q,
  \label{eq:cr_transport}
\end{equation}

\noindent
where $e_\mathrm{cr}$ is the CR energy density, $\hat{K}$ is the diffusion tensor, $\vec{v}$ is the plasma velocity, $p_\mathrm{cr}$ is the CR pressure and $Q$ is the source term of $e_\mathrm{cr}$. CR pressure is related to the $e_\mathrm{cr}$ via the adiabatic CR index:

\begin{equation}
  p_\mathrm{cr} \equiv (\gamma_\mathrm{cr} - 1) e_\mathrm{cr}.
  \label{eq:cr_pressure}
\end{equation}

\noindent
The diffusion of CRs is anisotropic \citep{giacalone99} with respect to the magnetic field, hence the diffusion tensor $\hat{K}$ from Eq.~\ref{eq:cr_transport} is defined as follows \citep{ryu03}:

\begin{equation}
  K_{ij} = K_\perp \delta_{ij} + (K_\parallel - K_\perp) n_i n_j,
\end{equation}

\noindent
where $K_\perp$ and $K_\parallel$ are the diffusion coefficients perpendicular and parallel to the local magnetic field, respectively, and $n_i \equiv B_i/B$ is the $i$-th component of the vector tangent to the local magnetic field.

In the CR-driven dynamo model the source term $Q$ in Eq.~\ref{eq:cr_transport} is attributed to the SN remnants \citep{hanasz04} in which CR particles are accelerated. A fraction of the kinetic energy output of the SN explosion is converted into the acceleration of the CRs in the shock front.  The conversion rate is 10-30\% \citep{dorfi00} of the kinetic energy. We assume that the kinetic energy output of the SN explosion is $10^{51}$~erg and the conversion rate is 10\%.  Each SN explosion is modelled by a 3D Gaussian distribution of CR energy input and is added to the source term~$Q$ in Eq.~\ref{eq:cr_transport}. Some of the explosions are magnetized and besides the input of CR energy, a randomly oriented dipole magnetic field is injected into the ISM\@.  The magnetized SN explosions occurs only for the first $t_\textrm{mag}$ time and only 10\% of all explosions in that time are magnetized. This is to seed the dynamo action.

The position of each SN explosion is chosen randomly according to the local gas density.  \citet{schmidt59} and \citet{kennicutt89} showed that in a simple self-gravitational picture the large-scale SFR volume density $\rho_\textrm{SRF}$ scales with gas density as $\rho_\textrm{SFR} \propto \rho^{3/2}$.  Using this relation we can build a cumulative distribution function of SN explosions in a galactic disk. Assuming a constant SN explosion frequency, $f_\textrm{SN}$, we can calculate a number of exploding stars and find their positions according to a cumulative distribution function in each time step.  The $f_\textrm{SN}$ is modulated in time with a period of 200\,Myr, in which for the first 40\,Myr  the SNe are active, before being set to zero;  after the following 160\,Myr the process is repeated. The $f_\textrm{SN}$ value represents the SN explosion frequency during the active interval, so to compute the mean it should be multiplied by 40\,Myr/200\,Myr = 0.2. The period 200\,Myr has been found by \cite{siejkowski10} as the most effective for the amplification of the magnetic field in dwarf galaxies. Additionally they found that the longer the period of silence in SN activity the more effective the amplification.

The dwarf/irregular galaxy gravitational potential well is given by two components: dark matter (DM) halo and thin stellar disc.  This type of galaxy does not have bulge \citep{governato10} which is usually present in bigger objects, like grand-design spirals.  The stars are distributed in infinitesimally thin Kuzmin's disk which generates a potential well of the form \citep[e.g.][]{gal_dyn}:

\begin{equation}
  \phi_*(R,z) = - \frac{G M_*}{\sqrt{R^2 + (a + |z|)^2}},
\end{equation}

\noindent
where $G$ is the gravitational constant, $M_*$ is the total mass of stars, $a$ is the radial scalelength and $R$ is the radius in equatorial plane ($z=0$), that is, $R \equiv \sqrt{x^2 + y^2}$.

The DM halo is approximated by a ``modified isothermal sphere'' which belongs to a class of double (broken) power-law density distributions \citep[see][]{gal_for_evo}. The density profile is given by $\rho_{\mathrm{h}}^{\mathrm{iso}}(r) = \rho_{0}/[1 + (r / r_0)^2]$, and the potential is described as:

\begin{equation}
  \phi_{\mathrm{h}}^{\mathrm{iso}}(r) = 4\pi G \rho_{0} r_0^2 \left[ \frac{1}{2}\ln{(1+x^2)} + \frac{\arctan{x}}{x} \right],
  \label{eq:isosphere}
\end{equation}

\noindent
where $\rho_{0}$ is the central density, $r_0$ is the core radius and $r$ is the distance to the centre $r \equiv \sqrt{R^2 + z^2}$, and $x \equiv r/r_0$.

The initial condition of the galaxy model is set to hydrostatic equilibrium. The gas density distribution in the equatorial plane is assumed to have following form:

\begin{equation}
  \rho(R,z=0)=\frac{\rho_g}{\left[1 + \left(\frac{R}{R_c}\right)^2\right]^2},
\end{equation}

\noindent
where $\rho_g$ and $R_C$ are the central gas density and core radius, respectively. Then to find the global gas distribution we use the "potential method" following \cite{wang10}. From the gas distribution the CR component distribution is found assuming that they are in pressure equilibrium, namely $\beta \equiv p_\textrm{CR}/p_\textrm{gas} = 1$.  The magnetic field in $t = 0$ is set to zero ($\alpha \equiv p_\textrm{mag}/p_\textrm{gas} = 0$).

\subsection{Model description}

We setup the gas temperature to about 6000\,K, which corresponds to $c_s = 7$\,km\,s$^{-1}$. The magnetic diffusivity is set to $3\times10^{25}$\,cm$^2$\,s$^{-1}$ following \cite{hanasz09param}, who have shown that this value is optimal for the growth of the magnetic field in the buoyancy-driven dynamo. Following \cite{ryu03} the CR adiabatic index was set to $\gamma_\textrm{cr} = 14/9$, and the CR diffusion coefficients are following: $K_\perp = 3 \times 10^{26}$\,cm$^2$\,s$^{-1}$ and $K_\parallel = 3 \times 10^{27}$\,cm$^2$\,s$^{-1}$. \cite{strong07} argues that the typical value for CR diffusion coefficient is about $(3\div5) \times 10^{28}$\,cm$^2$\,s$^{-1}$, however this highly impacts the time-step value and makes the simulation unfeasible. This issue has been investigated by \cite{hanasz09param}, who have shown that the magnetic field growth only slightly depends on the $K_\parallel$ value, but the key factor is the anisotropy of the CR diffusion that we apply. The 3D space domain is represented as a grid of 256$\times$256$\times$128 points in $x$, $y,$ and $z$ direction, respectively. We set the boundary condition to outflow in all directions.

We define two reference models with rotational velocities of approximately 50\,\kms and 
100\,\kms. The parameters of the models are given in Table~\ref{tab:basic_models}. 
We set $f_\textrm{SN}$ to $3 \times 10^3$\,kpc$^{-2}$\,Gyr$^{-1}$ for the slower 
rotation model, and three times lower for the faster rotation model.
For both models such values give similar global SFR$\approx 0.025$\,\Msun\,yr$^{-1}$  
in the ongoing phase of the SN activity (Sect. 2.1) and a mean value of 
the global SFR of about 0.005\,\Msun\,yr$^{-1}$ during the 200\,Myr activity period.
These values of parameters also provide stable code execution during the starburst 
phase of galaxy evolution. The corresponding actual galaxies can be found among 
non-starbursting dwarf galaxies \citep{chyzy11}; for example IC\,1613 with a global 
SFR of 0.003\,\Msun\,yr$^{-1}$.

\begin{table}
\centering
\caption{Parameters of the reference models.}
\label{tab:basic_models}
\begin{tabular}{lccl}
\hline\hline
Model name & v50 & v100 & \\
\hline
Parameter & & & Unit \\
\hline
$M_*$     & 2.4  & 15   & $10^9$\,\Msun \\
$\rho_0$  & 10.8 & 7.0  & $10^6$\,\Msun\,kpc$^{-3}$ \\
$r_0$     & 1.8  & 2.2  & kpc \\
$r_g$     & 29.5 & 29.5 & $10^6$\,\Msun\,kpc$^{-3}$ \\
$R_c$     & 1.0  & 2.0  & kpc \\
$f_\textrm{SN}$ & $3\times10^3$ & $1 \times 10^3$ & kpc$^{-2}$\,Gyr$^{-1}$ \\
galaxy radius & 4.5 & 7.5 & kpc \\
grid size & 47 & 78 & pc \\
\hline
Rotation speed & 50 & 100 & \kms \\
\hline
\end{tabular}
\end{table}

To study the influence of the starburst events on the magnetic field evolution in the galaxy we introduce such events into the reference models. This is modelled by a single event (one period of SN modulation) during which we inject $k \times f_\textrm{SN}$. For model v50 this event occurs at $t=6$\,Gyr and we investigate $k = 2, 5$ and for model v100 at $t=4$\,Gyr and we investigate $k=2$. 
The event lasts for 100\,Myr during the 200\,Myr period, and the mean SN activity is $0.5 k \times f_\textrm{SN}$.
Additionally we modify this event to inject the SNe only in one quarter of the galaxy, to study the localised burst events.
We note that the effective $f_\textrm{SN}$ value in the case of a localised event is four times stronger in the quarter. The total number of SN explosions is the same for the reference models and the localised ones.
The base levels in the models and the starburst events are shown in Table~\ref{tab:sfr}.

\begin{table}
\centering
\caption{The applied $f_\textrm{SN}$ in $10^3\,\textrm{kpc}^{-2}\,\textrm{Gyr}^{-1}$ for our models. The base level is used throughout the whole simulation, except if there is a starburst event, where the SFR is enhanced. The region describes whether the starburst event is in the whole disk or localized only to one quarter of the galaxy.}
\label{tab:sfr}
\begin{tabular}{lccc}
\hline
\hline
Model & Base level & Starburst event  & Region \\
\hline
v50     & 3.0 & --   & -- \\
v50x1Q  & 3.0 & 3.0  & quarter \\
v50x2   & 3.0 & 6.0  & whole disk \\
v50x2Q  & 3.0 & 6.0  & quarter\\
v50x5   & 3.0 & 15.0 & whole disk \\
v50x5Q  & 3.0 & 15.0 & quarter \\
\hline
v100    & 1.0 & --  & -- \\
v100x1Q & 1.0 & 1.0 & quarter \\
v100x2  & 1.0 & 2.0 & whole disk \\
v100x2Q & 1.0 & 2.0 & quarter \\
\hline
\end{tabular}
\end{table}

Throughout the paper we use the following notation for the model names: ``v'' and a number stands for the rotational velocity, ``x'' and a number denotes the burst event and the number corresponds to the $k$ value, ``Q'' marks models in which the event occurs only in a quarter of the galaxy disc. For example, the name v50x2Q represents a model of a galaxy with a rotation as fast as 50\,\kms, in which the burst event is two times the base $f_\textrm{SN}$ value and the SNe has been injected only in a quarter of the galactic disc.

\section{Results}
\begin{figure}
\centering
\includegraphics[width=0.49\textwidth]{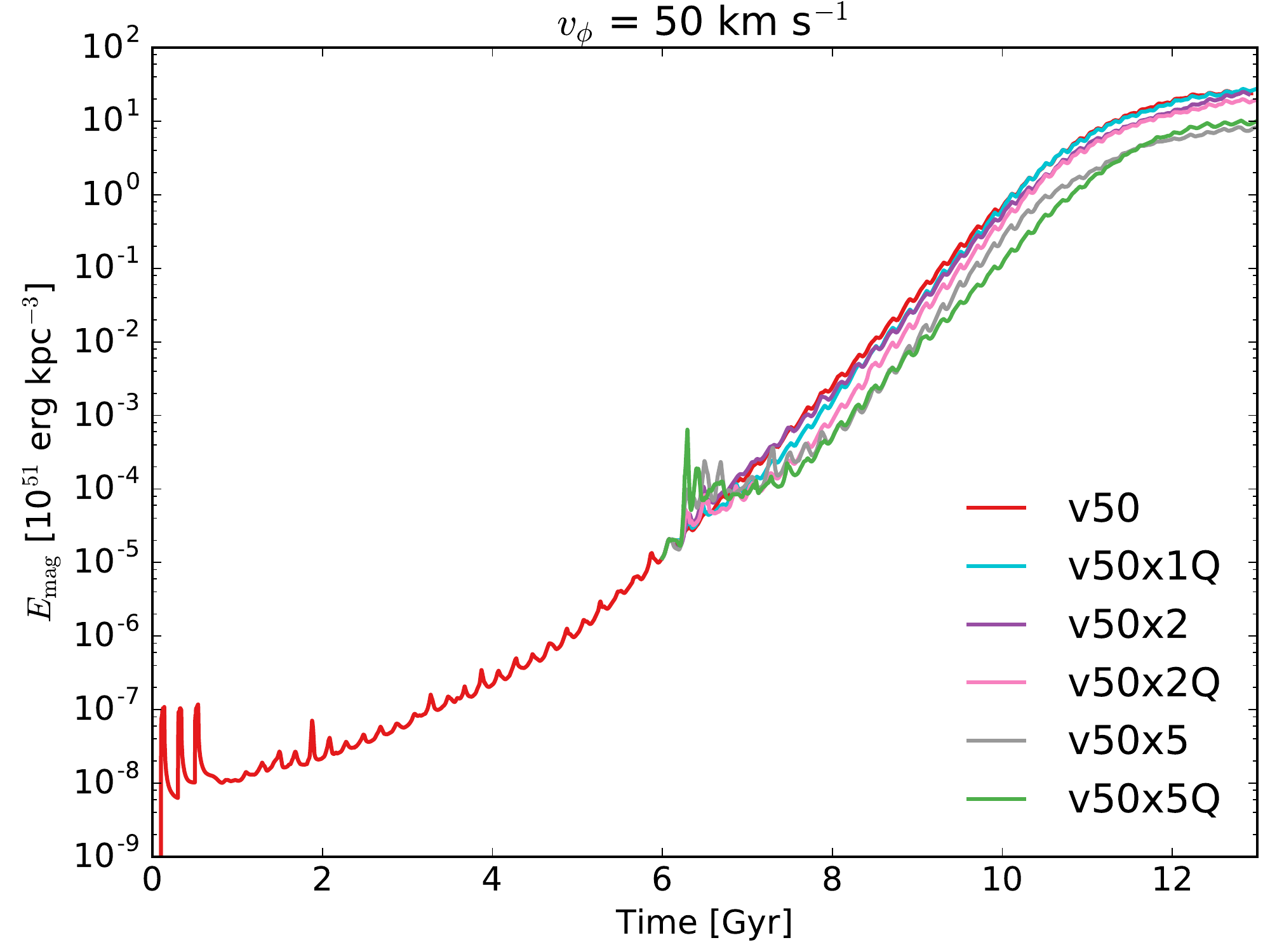}
\includegraphics[width=0.49\textwidth]{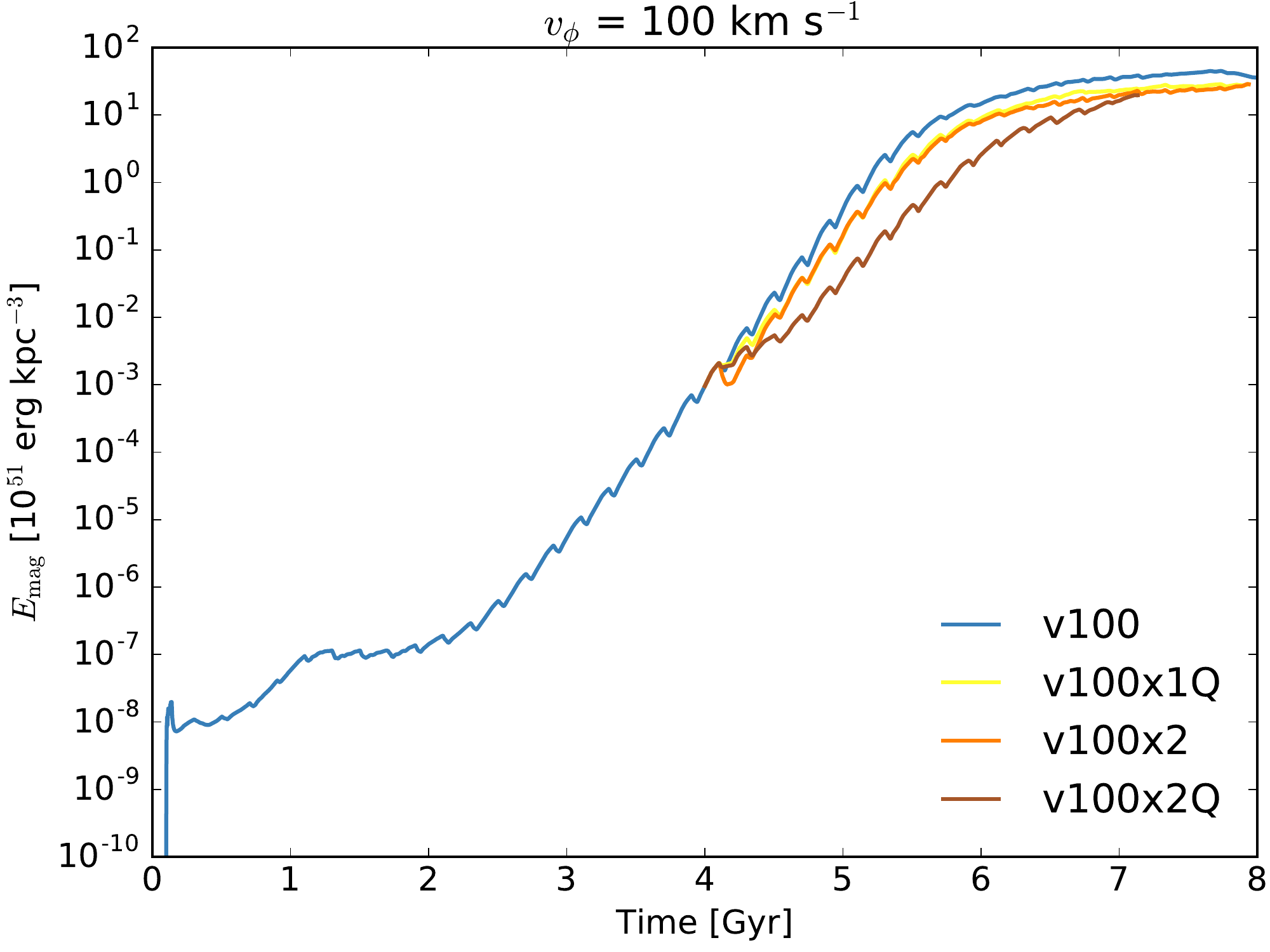}
\caption{Evolution of the total magnetic energy density.}
\label{fig:b2_evolution}
\end{figure}

\begin{figure}
\centering
\includegraphics[width=0.49\textwidth]{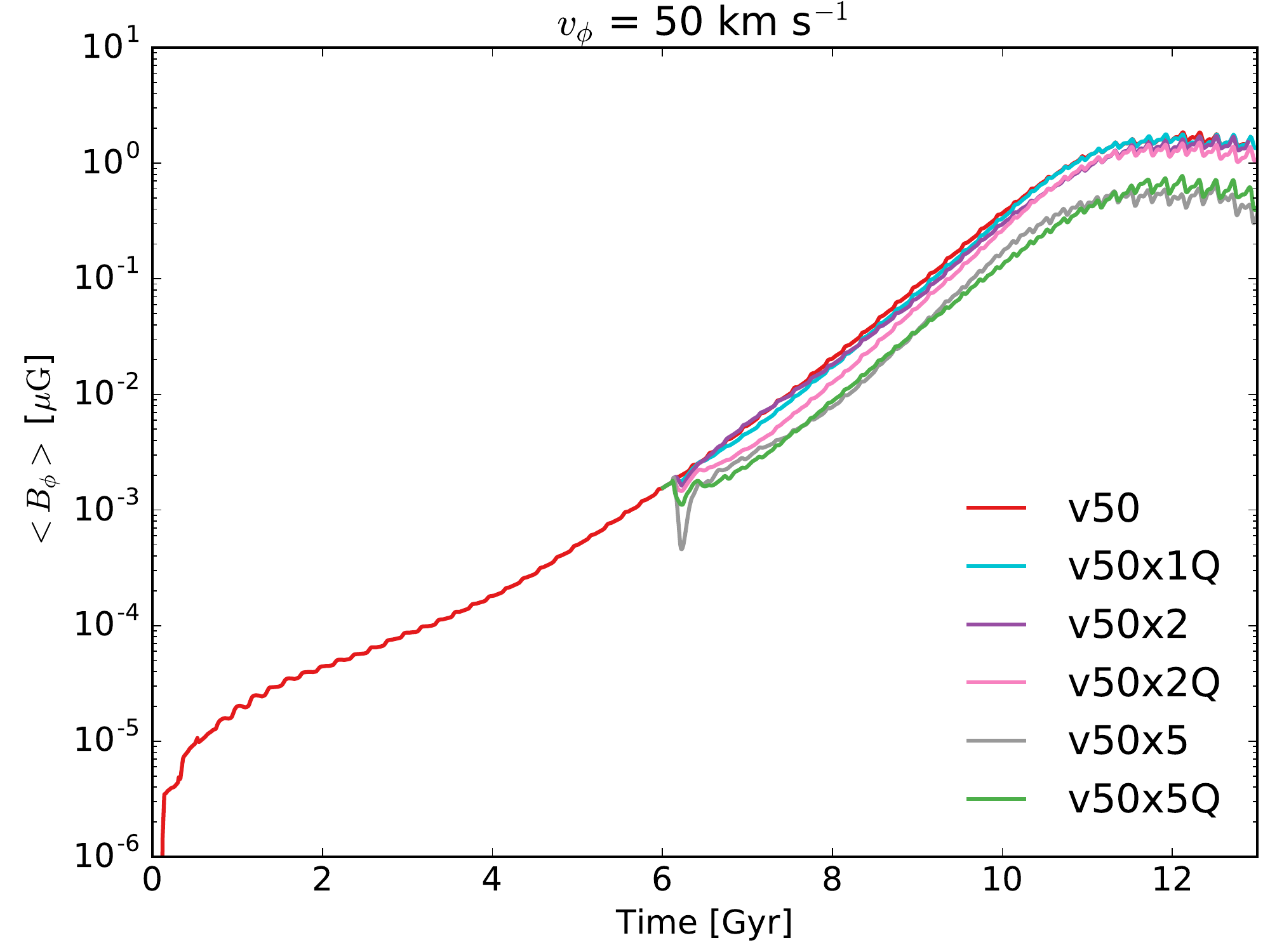}
\includegraphics[width=0.49\textwidth]{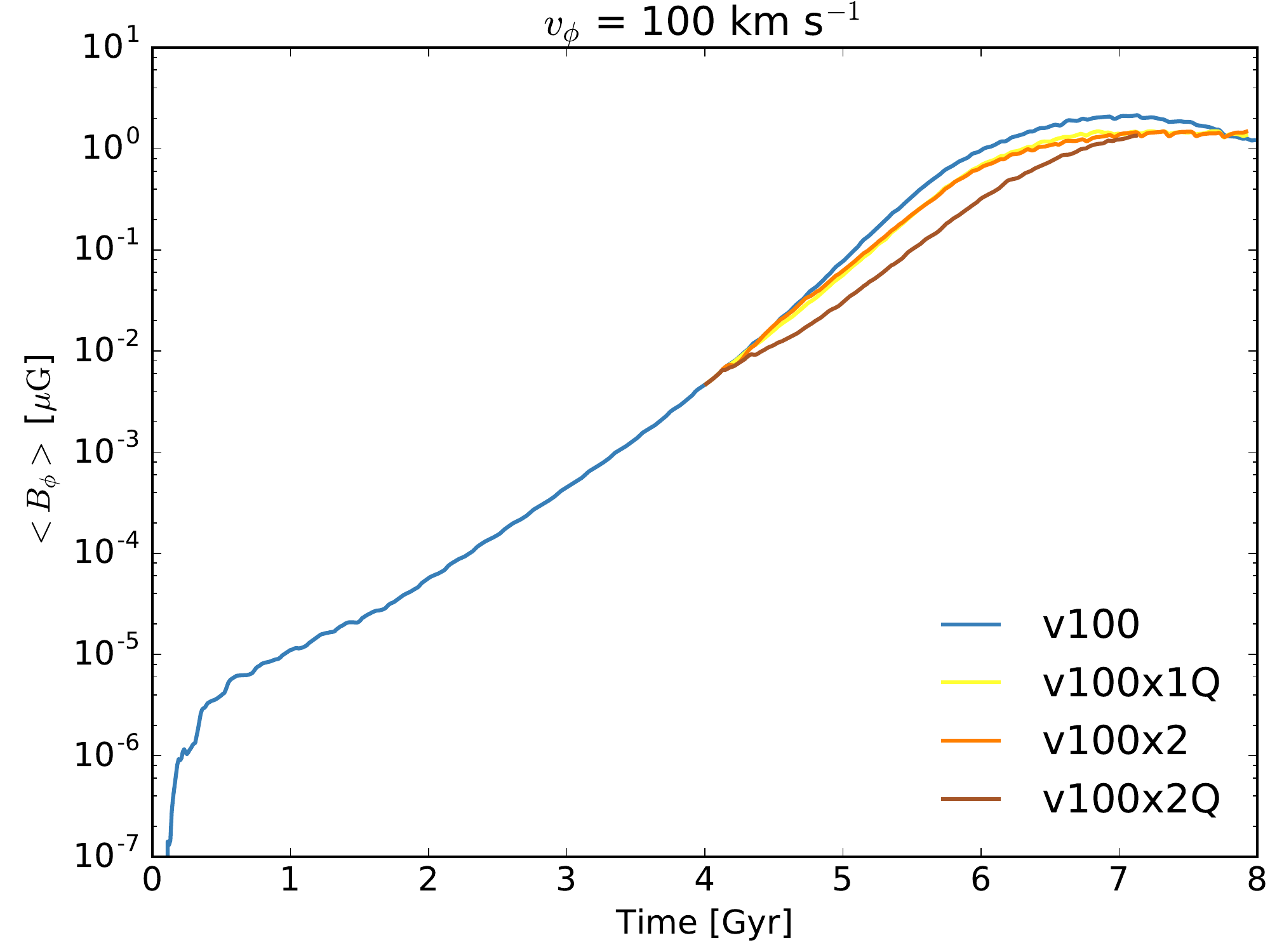}
\caption{Evolution of the mean azimuthal magnetic component. To use the logarithmic scale for the azimuthal component plot the absolute value is plotted;  if the flux starts rapidly dropping to very low values and then rapidly grows, this is a signature of sign turnover.}
\label{fig:bfi_evolution}
\end{figure}

The evolution of the total magnetic energy, $E_\textrm{mag}$ , and mean magnetic azimuthal component, \bfi, are shown in Figs.~\ref{fig:b2_evolution} and \ref{fig:bfi_evolution}, respectively.
To estimate the \bfi we calculate the $B_\phi^{i,j,k}$ in each grid point and then we sum them all and divide by the number of grid points.
We restrict the summing of the grid points to $|z|\leq1$\,kpc.

All the models investigated show an exponential growth of the magnetic field in time. In the case of models with $v_\phi^\mathrm{max}=50$\,\kms in the beginning of the simulation we inject three events of magnetized SN explosions (Fig.~\ref{fig:b2_evolution}). This is visible in the magnetic energy plot as three peaks in the first 500\,Myr. Then the magnetic field injection is stopped, and the growth is caused only by the dynamo operation. We found that for models with $v_\phi^\mathrm{max}=100$\,\kms only one such event is enough to seed the dynamo.

\begin{figure}
\centering
\includegraphics[width=0.49\textwidth]{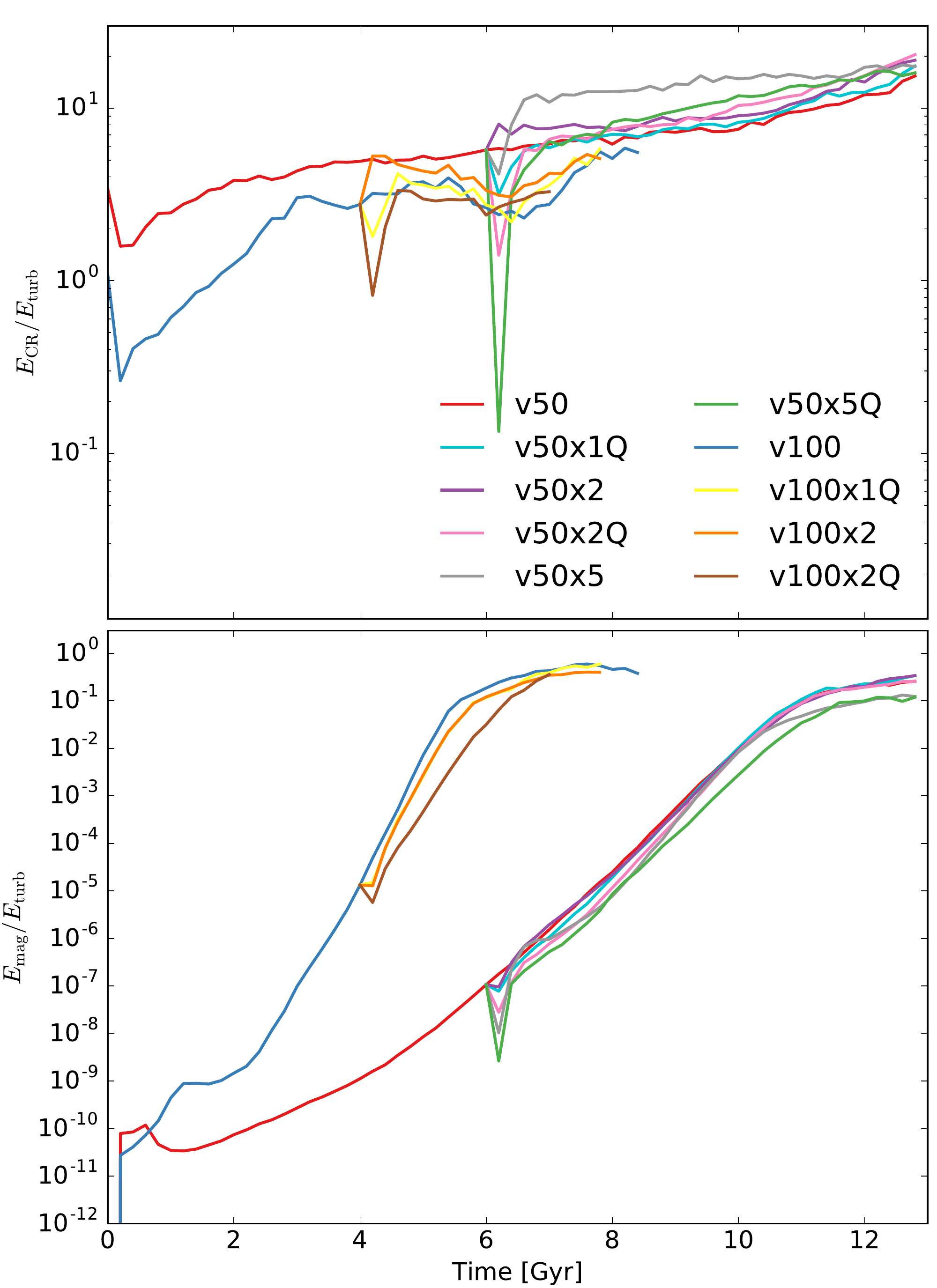}
\caption{Evolution of the energy ratios. The upper panel shows $\alpha_E$ and the lower one  $\beta_E$.}
\label{fig:eq}
\end{figure}

The models with $v_\phi^\mathrm{max}=50$\,\kms show constant growth of the magnetic field energy, just after the injection of magnetized SNe is stopped. The $e$-folding time is about 0.5\,Gyr for the magnetic energy (see Table~\ref{tab:efolding}) and the saturation point is reached after 12\,Gyr. The azimuthal component evolution shows two phases: the first, the slow one which occurs between $t=1$\,Gyr and 4\,Gyr, and the second, the fast one after $t=4$\,Gyr which lasts until the saturation point around $t=12$\,Gyr. The mean $e$-folding time for \bfi is 0.91\,Gyr.

The faster rotators, after first injection of the magnetic field, start to amplify the magnetic field energy exponentially and around $t=1$\,Gyr the growth stops for the next 1\,Gyr. After that the dynamo restarts, and the amplification is exponential with $e$-folding as high as 0.25\,Gyr (see Table \ref{tab:efolding}). The saturation is reached around $t=5.5$\,Gyr. The mean azimuthal component of the magnetic field is amplified exponentially with the $e$-folding time equal to 0.44\,Gyr. 

We show the $e$-folding times of magnetic field evolution for all the models investigated in Table~\ref{tab:efolding}. Since the burst episodes influence the magnetic evolution we report the global mean values for $1\,\textrm{Gyr}<t<11\,\textrm{Gyr}$ and $0.5\,\textrm{Gyr}<t<5.5\,\textrm{Gyr}$ for v50 and v100, respectively. For all models representing galaxies with rotation of about 50\,\kms we obtain the $e$-folding times between $8\,\textrm{Gyr}<t<11\,\textrm{Gyr}$, and for 100\,\kms between $4.5\,\textrm{Gyr}<t<5.5\,\textrm{Gyr}$.

\subsection{Energy balance}
In our models of galaxies there are three components which contribute to the total energy of the system, namely\ thermal gas, CRs, and magnetic fields. To show the energy evolution for each component, the following ratios are calculated:

\begin{equation}
  \alpha_E \equiv E_\textrm{mag}/E_\textrm{trub}\quad\textrm{and}\quad \beta_E \equiv E_\textsc{CR}/E_\textrm{trub}.
\end{equation}

\noindent
The turbulence energy, $E_\textrm{turb}$ is estimated  by:
\begin{equation}
    E_\mathrm{turb}=\frac{1}{2} M (\mathrm{Var}(v_r)+\mathrm{Var}(v_\phi)+\mathrm{Var}(v_z)),
\end{equation}
\noindent
where $v_r$, $v_\phi$, and $v_z$ are the velocity vector components expressed in aziumthal coordinates, $M$ is the total gas mass in the domain, and $\mathrm{Var}(X)$ is the variance, that is, $\mathrm{Var}(X) = \Sigma_{i=1}^{n}(x_i - \mu)^2/n$, where $\mu = \Sigma_{i=1}^{n}{x_i}/n$.

The evolution of $\alpha_E$ and $\beta_E$ for the reference models are shown in Fig.~\ref{fig:eq}.
We restrict the computations of $\alpha_E$ and $\beta_E$ to the disk region, that is, $|z|<1$~kpc.
The energy density of the CRs during the evolution is slowly increasing because of the SNe injection. The final value of $\beta_E$ is between 3.0 and 20.0 for each of the investigated models.
The evolution of $\alpha_E$ ratio, which is crucial for the dynamo action is shown as well. The initial value, at $t=0$ is set to 0, and after the simulation starts, the magnetic field is seeded by the magnetized SN explosions. The energy ratio $\alpha_E$ after first injections is between $10^{-11}$ and $10^{-10}$ and depends on the galaxy size. The exponential growth is then visible until the saturation around $\alpha_E \simeq 0.5$.

The issue of the energy balance in the system between turbulent kinetic energy, CRs, and magnetic fields
has been investigated by \cite{hanasz09param} in the local simulations of CR-driven dynamo
in massive galaxies. They found that the evolution of $E_\textsc{cr}/E_\textrm{trub}$ depends on the
CR diffusion coefficients and if the latter is higher, the ratio after the simulation is lower. In
our work the adopted value of the CR diffusion is 10\% of the realistic values because of the
time step limitations. The explicit algorithm of the CR diffusion causes the time step to be
significantly short when realistic values of diffusion coefficients are used. The simulation
results show that in the case of global dwarf galaxy simulation the CR energy is comparable to the turbulent kinetic energy.
\cite{hanasz09param} and \cite{snodin06} find excess CR energy in the numerical simulations of galaxy evolution and CR gas.
It also seems that the ratio of CR to other forms of energy in the
ISM is not yet well constrained by the observations \citep{strong07,stepanov12}.

\begin{table}

\caption{The $e$-folding times given in Gyr for the evolution of the magnetic energy (1) and azimuthal component of the magnetic field (2). The models denoted with a star show the global mean values, as defined in the text.}
\label{tab:efolding}

\centering
\begin{tabular}{lcc}
\hline \hline
Model & (1) & (2) \\
\hline

v50*    & 0.50 & 0.91 \\ 
v50     & 0.38 & 0.75 \\ 
v50x1Q  & 0.36 & 0.71 \\
v50x2   & 0.38 & 0.74 \\
v50x2Q  & 0.35 & 0.70 \\
v50x5   & 0.36 & 0.74 \\
v50x5Q  & 0.38 & 0.77 \\
\hline

v100*   & 0.25 & 0.44 \\ 
v100    & 0.18 & 0.34 \\
v100x1Q & 0.19 & 0.38 \\
v100x2  & 0.19 & 0.40 \\
v100x2Q & 0.23 & 0.46 \\

\hline

\end{tabular}
\end{table}
 
\section{Discussion and conclusions}
Our models represent dwarf and Magellanic-type irregulars. Both reference 
models (v50 and v100) correspond to a calm phase of galactic evolution.
Our mean surface density of the SFR ($f_\mathrm{SN}$) covers
the lower end of surface star formation rate (SSFR).  The objects IC\,1613 and IC\,2574 may represent good real examples
\citep[see e.g. Fig.\,5. in][]{jurusik}.
Our results show that a faster rotator amplifies the magnetic field
much faster, regardless of the lower star formation activity applied.
Stronger action of the dynamo for a faster rotating galaxy was also visible in our
previous simulations of a small-mass dwarf galaxy \citep{siejkowski14}.

Our main goal was to study the consequences of the burst of star formation 
in the history of the galaxy. These effects have not been simulated before.
In models v50x2, v50x5, and v100x2 we 
applied an increased SFR once in the galactic evolution.
Slower rotators (v50x2 and v50x5) show rapid increase of the magnetic field
as an immediate response to the increased SFR. However, after the initial
growth, when the burst of star formation stops, the magnetic field value
returns to its level corresponding to the unperturbed galactic evolution, as 
seen in Fig. \ref{fig:bfi_evolution}.
Total magnetic energy calculated in the whole simulation domain even falls below
its ``unperturbed'' level, clearly showing the smaller growth rate
(Fig. \ref{fig:b2_evolution}) over about 1\,Gyr until the dynamo process comes
to the same activity rate as before the burst of star formation.
More violent starburst (model v50x5) events lead to a lower saturation
level of the magnetic field value at the end of the simulated galactic history.

Our faster rotator (model v100x2) evolution does not show the initial increase
of magnetic field as a response to the starburst event. The visible amplification rate, as well as its saturation level, are only 
slightly lower.

Star formation bursts frequently occur only in some parts of a galaxy (e.g.
IC\,10, LMC, NGC\,4449). We tried to simulate such a situation by restricting
the increased SFR to only one quarter of the galactic disk (models v50x1Q,
v50x2Q, v50x5Q, v100x1Q, and v100x2Q). The magnetic field evolution 
caused by such localised starburst episodes is similar
to in the corresponding ``non-Q'' models. All effects described above are 
present, but emphasised. This can be explained, knowing that while restricting 
the area for a burst of star formation, we kept the total number of SN
explosions the same, that is, the local SFR in the burst area is four times higher.

\begin{figure*}
\centering
\includegraphics[width=0.49\textwidth]{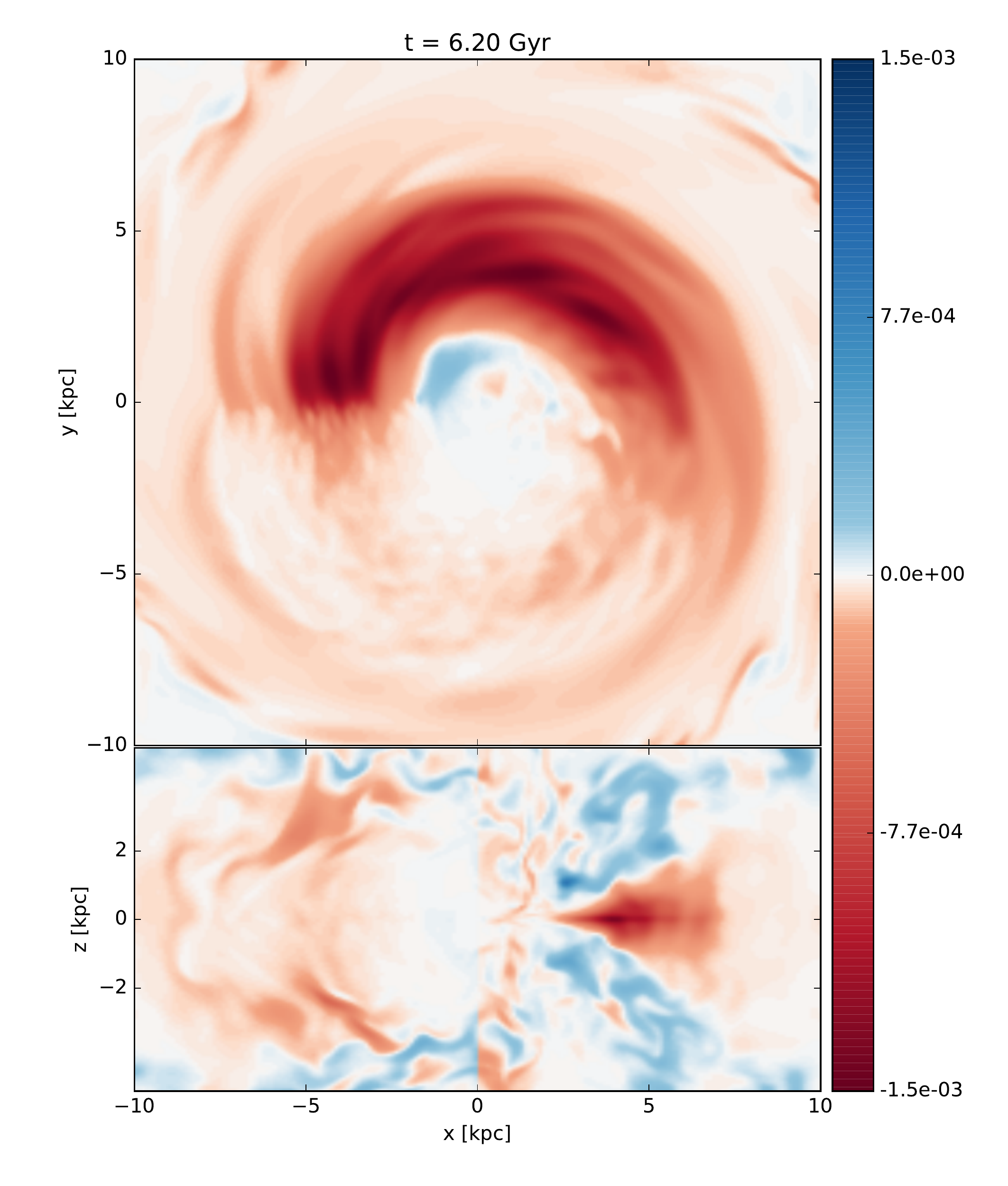}
\includegraphics[width=0.49\textwidth]{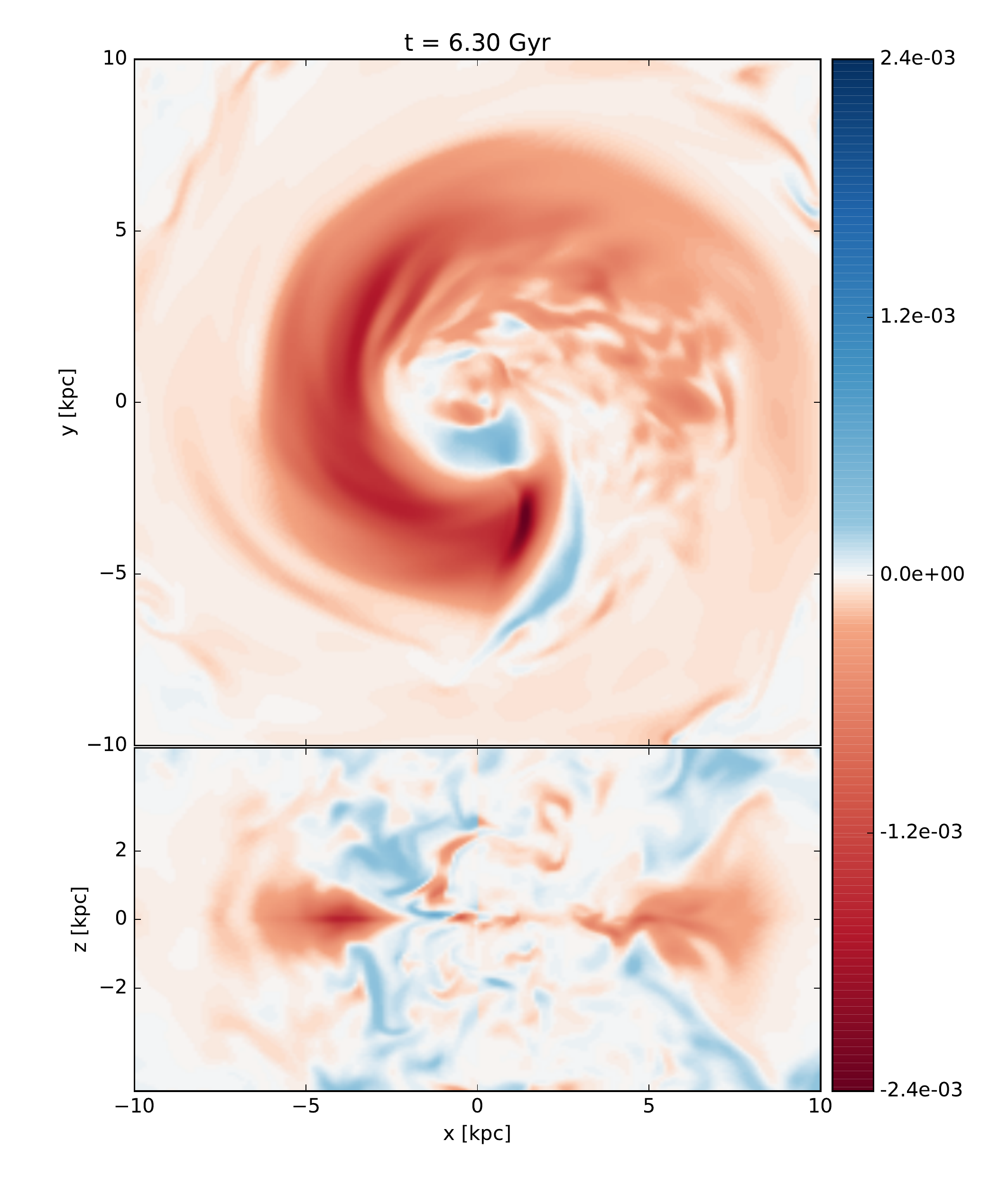}
\caption{Two snapshots of the azimuthal magnetic field component in the 
galactic disk and in the plane perpendicular to it. Left: at the beginning of a starburst event; Right: at the end of the starburst. We note the change of the
colour-scale in both plots.}
\label{fig:burst}
\end{figure*}

It is interesting to see what is responsible for the growth of the magnetic
field in the star formation burst event. Figure \ref{fig:burst} shows the 
azimuthal magnetic field component in the galactic plane and in the plane
perpendicular to it at the beginning (left panel) and at the end (right panel)
of the burst event. Initially the increased SN activity disrupts the regularity
of the magnetic field (its dominant azimuthal component). At the same time 
increased turbulence in the interstellar medium gives rise to more efficient 
dynamo action -- most efficient at the boundary between areas of highly turbulent
bursts and areas filled with more ordered magnetic field with no bursts.
The galactic differential rotation propagates the boundary over the whole 
disk giving rise to global increase of the dynamo efficiency in a short time.

\begin{figure}
\centering
\includegraphics[width=0.49\textwidth]{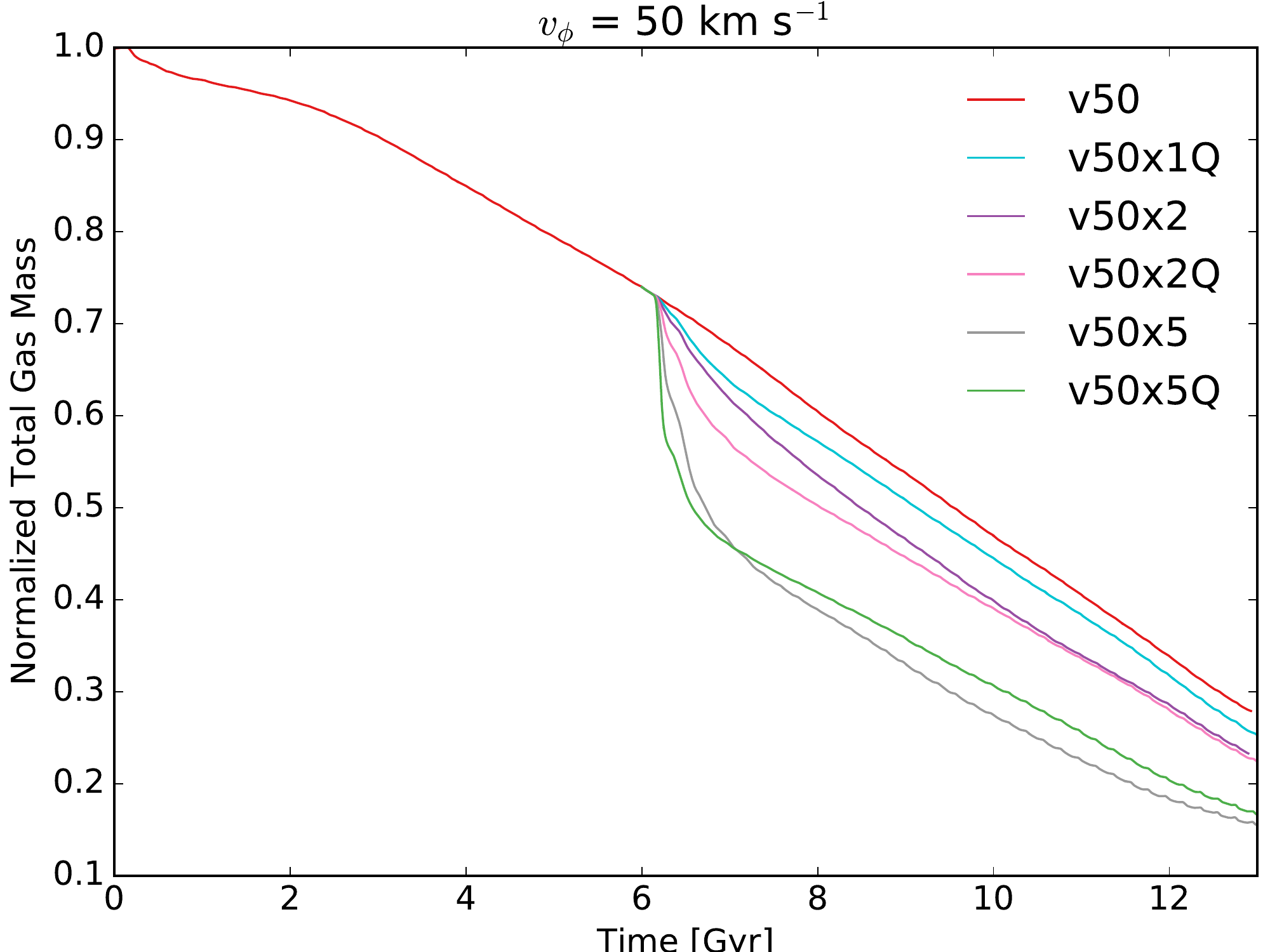}
\includegraphics[width=0.49\textwidth]{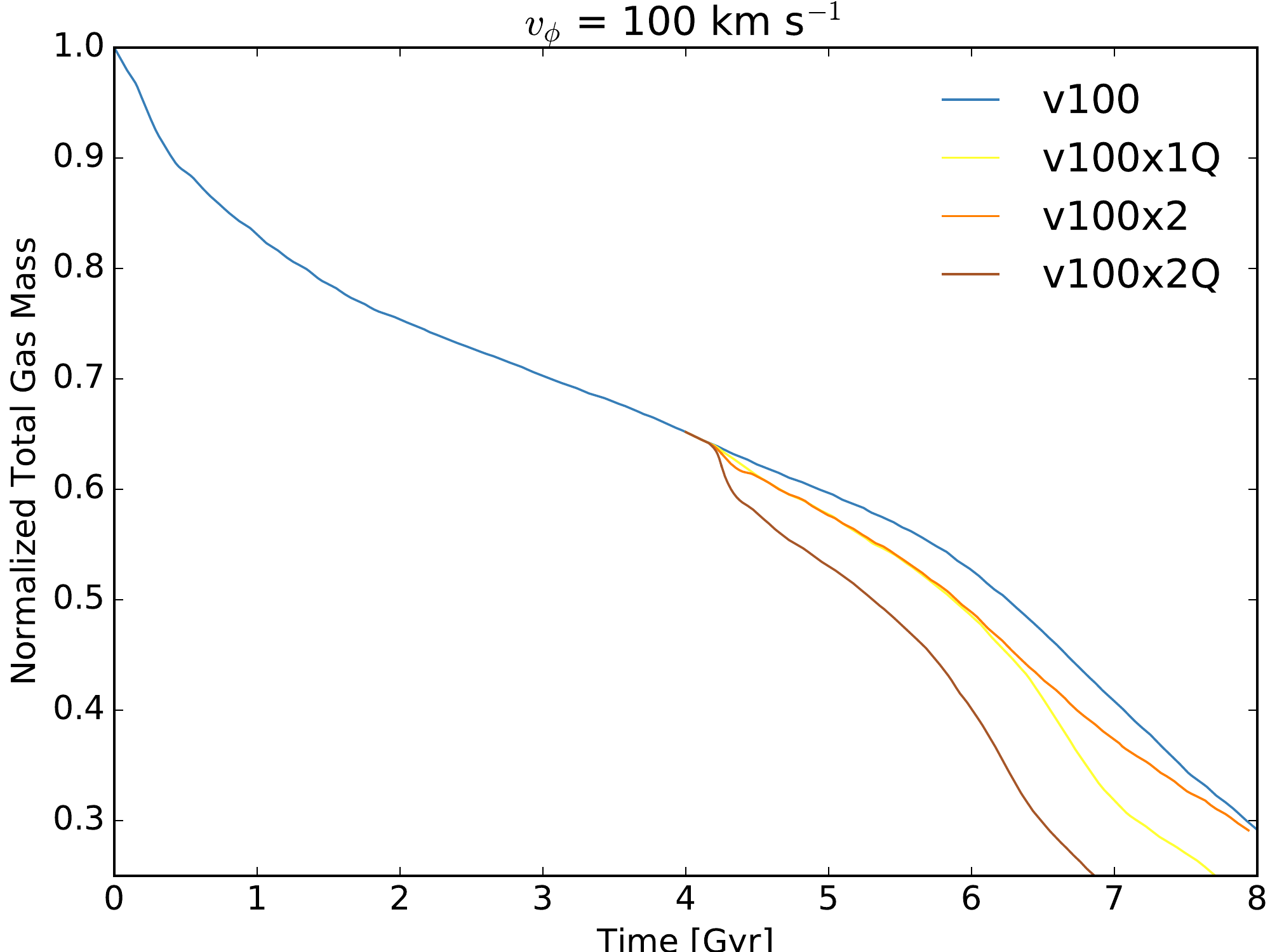}
\caption{Evolution of the total gas mass in a galaxy.}
\label{fig:mass_evolution}
\end{figure}

After the starburst event, the growth rate of the magnetic field decreases.
We suggest that this is the consequence of a large fraction of the 
magnetic field expelled out of the galactic body together with the interstellar
gas. The more violent the star formation involved is, the greater the volume of gas expelled 
from the galaxy (out of our computational domain), as shown in
Fig. \ref{fig:mass_evolution}. Thus the dynamo efficiency can still remain at 
the same level. The apparent smaller magnetic field growth-rate is 
only a consequence of increased galactic wind carrying the magnetic field
with it. A similar effect was observed in simulations of massive spirals
by Chamandy et al. \citep{chamandy} where stronger outflows in gaseous
arms locally suppress the dynamo action and lead to the formation of inter-arm
magnetic arms.
The magnetic field growth rate returns to the same value, as only the 
gas escape rate returns to the level before the burst event. This never
happens in the case of model v100x2Q where we observe an increased gas-escape rate 
until the end of the simulation.

Our simulations correspond to isolated galaxies. In actual galaxies,
the periods of enhanced star formation
could be triggered by tidal interactions with galaxy companions. This
could induce in turn the gas infall
\citep[e.g.][]{nidever}.  The loss of gas by galactic wind could
thus be supplemented with a new gas.
Tracing how such processes would affect the evolution of the magnetic field
in galaxies and change the above results requires new modelling.

From this work, we make the following conclusions.
CR-driven dynamos can effectively amplify the magnetic field, even under dwarf/Magellanic-type galaxy conditions. 
The magnetic fields are amplified with $e$-folding time for the regular component of about 0.91\,Gyr and reach the saturation level after 12\,Gyr for models with $v_\phi=50$\,\kms. For models where rotation is twice as fast, the $e$-folding is 0.44\,Gyr, and the saturation is reached after 6\,Gyr. The results are comparable to those reported by \cite{hanasz09global} and \cite{kulpa11}, who studied CR-driven dynamos in more massive galaxies and find $e$-folding times for the regular component of the magnetic field of around 0.3\,Gyr.
The studies of energy balance in the galaxy show that
the saturation point of the magnetic field growth is close to the equipartition level between the turbulent kinetic energy and the magnetic energy.
Turbulent kinetic energy and CR energy are not in balance due to the value of the CR diffusion coefficient being lower than in reality \citep[see e.g.][]{hanasz09param}.

We found that an episode of star-formation bursts increases the magnetic field,
but for a short time only. It is followed by a smaller amplification rate until
the dynamo action returns to the same efficiency rate as before the burst.
Localised star formation bursts emphasise the effects of the global 
starburst event, that is, the momentary growth of magnetic field (localised to the 
boundaries of a starburst), followed by smaller growth rate, and longer time
needed for return of the dynamo action to its previous efficiency. 
We noticed that in larger galaxies, which are more massive and rotate faster, the effects of the starburst event were stronger.
A substantial amount of gas is expelled from the
dwarf/Magellanic-type galaxy (up to about 70\% in 10\,Gyr). The starburst 
event allows galactic gas to be expelled even faster.

\begin{acknowledgements}
We thank the anonymous referee for detailed and constructive comments.
A substantial part of this work was supported by the Polish National Science Centre through grant 2012/07/B/ST9/04404.
This research was supported in part by PLGrid Infrastructure.
The plots presented in this paper are rendered using Matplotlib \citep{matplotlib}.
\end{acknowledgements}

\bibliographystyle{aa}
\bibliography{references}

\end{document}